\documentclass[twoside,twocolumn,9pt]{article}
\usepackage{extsizes}
\usepackage[super,sort&compress,comma]{natbib} 
\usepackage[left=1.5cm, right=1.5cm, top=1.785cm, bottom=2.0cm]{geometry}
\usepackage{balance}

\usepackage{sectsty}
\allsectionsfont{\scshape}

\usepackage{graphicx} 
\usepackage{lastpage}
\usepackage{float}
\usepackage{fancyhdr}
\usepackage{fnpos}
\usepackage[english]{babel}
\usepackage{array}
\usepackage[T1]{fontenc}
\usepackage[usenames,dvipsnames]{xcolor}
\usepackage{setspace}
\usepackage{hyperref}
\usepackage[squaren]{SIunits}

\usepackage{upgreek}
\usepackage{xspace}
\usepackage{placeins}
\usepackage{amsmath,amssymb}

\newcommand\hmin{\ensuremath{h_{\rm min}}\xspace}
\newcommand\hnot{\ensuremath{h_{\rm o}}\xspace}
\newcommand\hc{\ensuremath{h_\text{c}}\xspace}
\newcommand\vc{\ensuremath{v_\text{c}}\xspace}
\newcommand\cs{\ensuremath{c^*}\xspace}
\newcommand\etas{\ensuremath{\eta_\text{s}}\xspace}
\newcommand\etai{\ensuremath{[\eta]}\xspace}
\newcommand\strate{{\ensuremath{\dot\varepsilon}\xspace}}
\newcommand\stratec{\ensuremath{\dot\varepsilon_\text{c}}\xspace}
\newcommand\strater{\ensuremath{\dot\varepsilon_\text{R}}\xspace}
\newcommand\tc{\ensuremath{t_\text{c}}\xspace}
\newcommand\lr{\ensuremath{\lambda_\text{R}}\xspace}
\newcommand\lz{\ensuremath{\lambda_\text{z}}\xspace}

\newcommand\mw{\ensuremath{M_\text{w}}\xspace}
\newcommand\kb{\ensuremath{k_\text{B}}\xspace}
\newcommand\na{\ensuremath{N_\text{a}}\xspace}

\newcommand{\SR}[1]{\textcolor{black}{#1}}

\begin{document}

\twocolumn[
  \begin{@twocolumnfalse}
\vspace{3cm}

\begin{center}

    \noindent\huge{\textbf{\textsc{Transition to the viscoelastic regime in the thinning of polymer solutions}}} \\
    \vspace{1cm}

    \noindent\large{Sreeram Rajesh,\textit{$^{a}$} Virgile Thi\'evenaz,\textit{$^{a}$} and Alban Sauret\textit{$^{a}$}}$^{\ast}$ \\

    \vspace{5mm}
    \noindent\large{\today} \\

    \vspace{1cm}
    \textbf{\textsc{Abstract}}
    \vspace{2mm}

\end{center}

\noindent\normalsize{In this study, we investigate the transition between the Newtonian and the viscoelastic regimes during the pinch-off of droplets of dilute polymer solutions and discuss its link to the coil-stretch transition. The detachment of a drop from a nozzle is associated with the formation of a liquid neck that causes the divergence of the local stress in a vanishingly small region. If the liquid is a polymer solution, this increasing stress progressively unwinds the polymer chains, up to a point where the resulting increase in the viscosity slows down drastically the thinning. This threshold to a viscoelastic behavior corresponds to a macroscopic strain rate \stratec. In the present study, we characterize the variations of \stratec with respect to the polymer concentration and molar weight, to the solvent viscosity, and to the nozzle size, \textit{i.e.}, the weight of the drop. We provide empirical scaling laws for these variations. We also analyze the thinning dynamics at the transition and show that it follows a self-similar dynamics controlled by the time scale ${\stratec}^{-1}$. This characteristic time is different and always shorter than the relaxation time of the polymer.} \\

 \end{@twocolumnfalse} \vspace{0.6cm}

  ]

\makeatletter
\renewcommand*{\@makefnmark}{}
\footnotetext{\textit{$^{a}$~Department of Mechanical Engineering, University of California, Santa Barbara, California 93106, USA}}
\footnotetext{\textit{$^{*}$ asauret@ucsb.edu}}
\makeatother


\section{Introduction}

The formation of liquid droplets is a common phenomenon in everyday life~\cite{Villermaux2007, Villermaux2020, Kooij2018,raux2020spreading} in situations as diverse as sneezing,\cite{Scharfman2016, Mittal2020} spraying of pesticides,\cite{Makhnenko2021} inkjet printing,\cite{Derby2010, Lohse2021} bioprinting of tissues,\cite{Murphy2014} or combustion.\cite{Wei2015, Jaffe2015} When a liquid droplet detaches from a nozzle, the neck that binds the droplet to the nozzle thins down and eventually break. The break-up corresponds to a finite-time singularity.~\cite{Eggers1997} In many industrial applications, the liquid is not a homogeneous Newtonian fluid but often contains particles, solutes, cells, or other components. At certain length and time scales, these components interact with the flow and sometimes influence capillary flows dramatically. The instantaneous and localized change in rheology can even temporarily suppress the singularity.

In the case of polymers, the question of the time scale of the flow arises. \SR{At rest, a polymer chain adopts the shape of a coil to maximize its entropy, but under a strong extensional flow it stretches:\cite{smith1999single,nguyen2012flexible,Ingremeau2014}} this is the coil-stretch transition.~\cite{DeGennes1974} This conformational change modifies its interactions with the surrounding fluid. In the coiled state, the chain has the approximate shape of a sphere: only the monomers on the surface contribute to hydrodynamic friction. As the chain stretches, more monomers are exposed to the external flow, and the polymer exerts more friction onto the fluid. \SR{Therefore, the viscosity of a dilute polymer solution follows Einstein's law at low strain rates but increases dramatically at high strain rates.} Moreover, the chains can absorb some elastic energy so that the solution is viscoelastic.

\begin{figure*}[h]
\centering
  \includegraphics[width=0.99\textwidth]{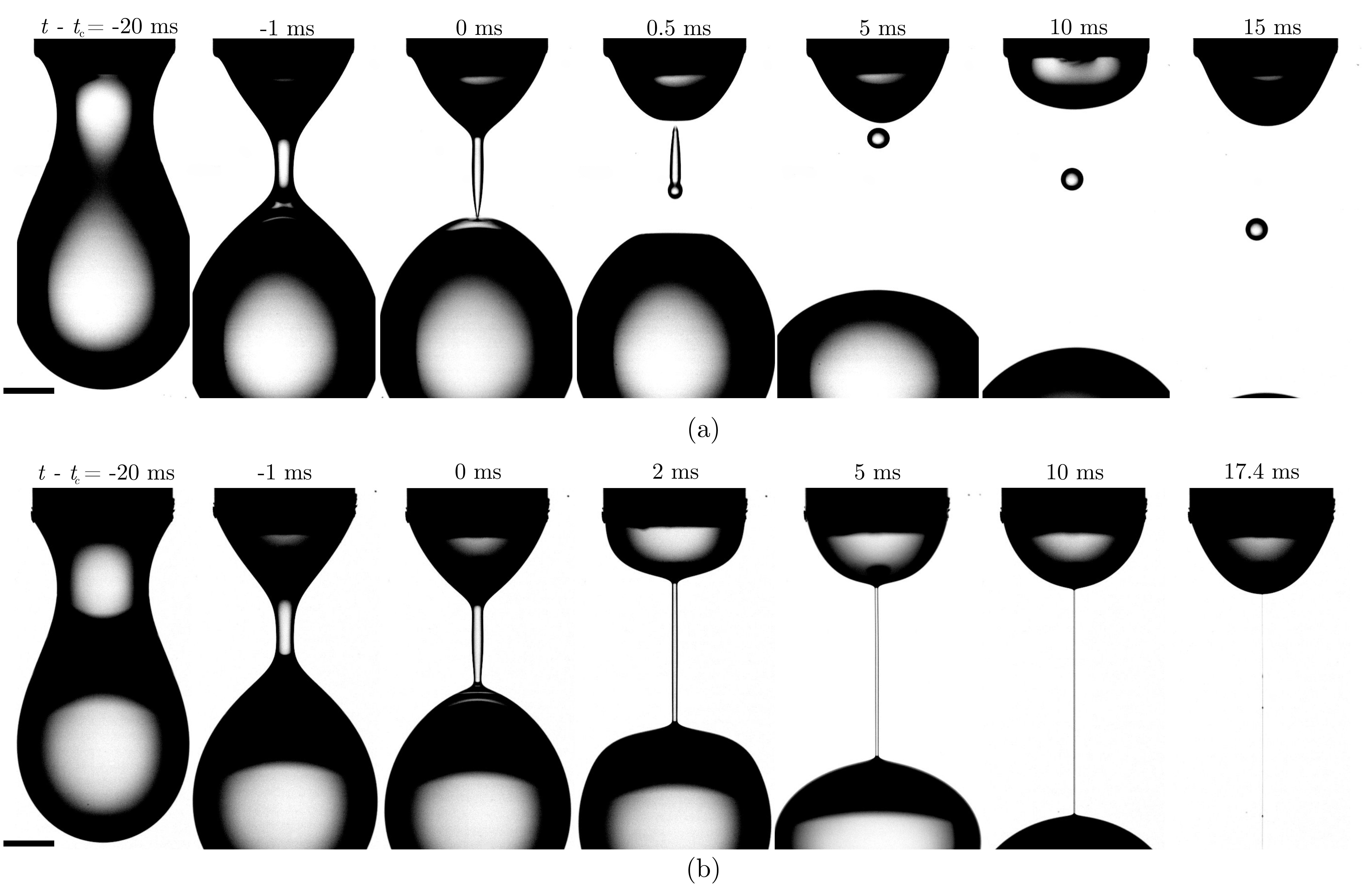}
  \caption{Pinch-off of a drop of (a) pure solvent (50/50 wt\% water-glycerol) and of
      (b) dilute polymer solution ($c=0.5\%$wt 300K PEO in 75/25 wt\% water-glycerol).
      Adding a small quantity of polymer prevents the singularity and the fast break-up of the liquid neck between the drop and the nozzle. The scale bars are 1 mm. The movies of these experiments are available in Supplemental Materials.
  }
  \label{fig:timeline}
\end{figure*}

The viscoelastic nature of a solution of polymers can be observed during the pinch-off of a drop. \SR{As shown in figure \ref{fig:timeline}, the evolution of the minimum thickness profile for a polymer solution [figure \ref{fig:timeline}(b)] is similar to that of a Newtonian drop [figure \ref{fig:timeline}(a)] in a certain interval ($-20 \,\milli\second < t-t_{\rm c} < 0 \,\milli \second $). But at some point, the neck turns into a long and slender filament.~\cite{Amarouchene2001,steinhaus2007dynamics,dinic2015extensional,dinic2017pinch,deblais2020self,jimenez2021rheologically} For $t>t_{\rm c}$, the thinning continues, but much more slowly than the initial necking, until the filament eventually breaks}. In the meantime, capillary instability may cause one or more large droplets to appear on the filament, the so-called ``beads-on-a-string'' instability.\cite{clasen2006beads,bhat2010formation} A secondary ``blistering'' instability, occurring in the last stages of thinning and forming many tiny droplets along the filament, has also recently been described.~\cite{Deblais2018}

Rather than considering the process of viscoelastic thinning, which has been characterized extensively in the past, the present study focuses on its onset. The sudden change in the thinning dynamics is due to the coil-stretch transition, as reported by Amarouchene~\textit{et al.}~\cite{Amarouchene2001} At rest, a polymer chain adopts the shape of a coil to maximize its entropy. Since stretching reduces entropy, it is opposed by a resisting force that increases linearly with the deformation $\varepsilon$.~\cite{de-gennes1979} If the force applied onto the chain is strong enough, the coil will unwind and stretch.~\cite{DeGennes1974}. \SR{This uncoiling has been observed experimentally for an isolated polymer,\cite{nguyen2012flexible} for polymer solutions in a microfluidic channel,\cite{Ingremeau2014} and numerically investigated for extensional flows of entangled polymer melts.\cite{Sefiddashti2018}} As soon as the external force becomes comparable to the Brownian motion, the resisting force $f$ becomes non-linear, $f \sim \varepsilon^{3/2}$, because the excluded volume hinders the stretching of the chain.~\cite{schroeder2018single} This was theoretically predicted in the 1970s by Pincus~\cite{pincus1976excluded} and only recently proven experimentally.\cite{saleh2009nonlinear} Once the polymer is sufficiently stretched, self-avoidance effects weaken, and the chain exhibits an ideal behavior that opposes a linear resisting force.~\cite{schroeder2018single} In other words, past a sufficient elongation, stretching the polymer chains suddenly becomes easier.

To investigate the transition between the Newtonian and the viscoelastic regime, we consider here the pinch-off of drops of dilute polymer solutions, with different polymer concentrations, molecular weight, and solvent viscosity. We present in section \ref{sec:2} our experimental methods and the polymers used. We then recall the main features of the thinning of a polymer solution in section \ref{sec:3}: first, the Newtonian regime, in which the solution thins down like an inviscid Newtonian liquid following a power law; then the viscoelastic regime, characterized by an exponential decay of the filament thickness. Existing results for the relaxation time are also recovered. We then describe the thinning in terms of strain rate and identify two critical quantities that mark the coil-stretch transition: the critical strain rate \stratec and the critical thickness \hc. We provide empirical scaling laws for the dependence of these parameters on the polymer concentration, solvent viscosity, and nozzle diameter. Finally, we show that the strain rate follows a self-similar dynamics around the transition, entirely controlled by the critical strain rate. This result enables us to highlight and discuss in section \ref{sec:4} the difference between two characteristic times, one associated with the coil-stretch transition and the other with the relaxation of the stretched polymer chains.

\section{Experimental methods} \label{sec:2}
\begin{figure*}[hbt]
\centering
  \includegraphics[width=0.96\textwidth]{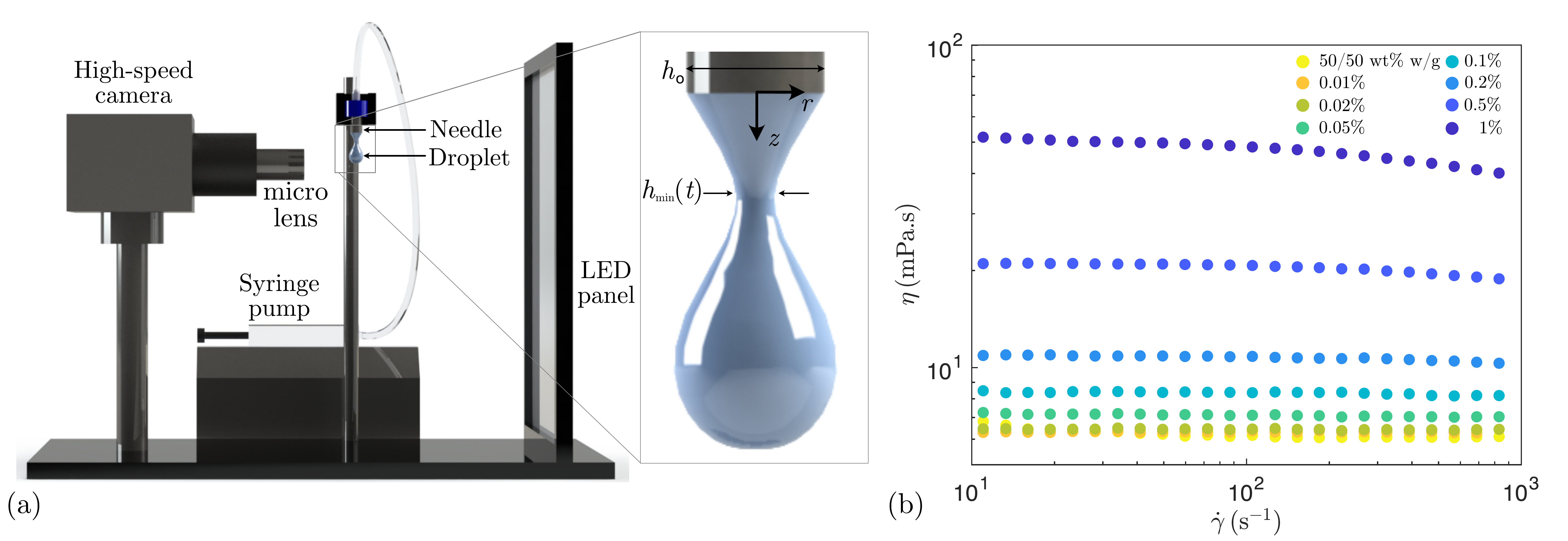}
  \caption{(a) Schematic of the experimental setup. \hmin corresponds to the minimal diameter of the liquid thread.
  (b) Evolution of the shear viscosity $\eta$ with the shear rate
  $\dot{\gamma}$ for polymer solutions prepared with different mass concentration $c$ of 300K PEO in a 50/50 wt\%  water-glycerol solvent.}
  \label{fig:setup}
\end{figure*}

  \begin{table*}
  \centering
 \small
     \begin{tabular*}{\textwidth}{@{\extracolsep{\fill}}|c|c|c|c|c|c|c|c|}
     \hline
     PEO $M_{\rm W} $ & Composition & c & [$\eta$] &  c/c* & $\eta_{\rm s}$ & $\eta$ & $\lambda_{\rm R}$\\
     (g/mol) &  (water/glycerol wt\%) & (wt\%) & $(m^3/kg)$ &  & (\milli\pascal\usk\second) & (\milli\pascal\usk\second) & (\milli\second)\\
     \hline
     $3\times10^5$ & 75/25 & 0.01-1 & 0.401 & 0.044-4.47 & 2.16 & 2.16-19.5 & 0.165-4.65\\
     $3\times10^5$ & 50/50 & 0.01-1 & 0.307 & 0.056-5.56 & 6.06 & 6.28-40.91 & 0.96-11.1 \\
     $4\times10^6$ & 75/25 & 0.01-1 & 2.65 & 0.36-37 & 2.16 & 2.56-98.3 & 24-505\\
      \hline
           \end{tabular*}
           \caption{Properties of polymer solutions used to carry out pinch-off experiments: $M_w$ is the molar weight of the polymer, $c$ the mass concentration in polymers, $\etai$ is the intrinsic viscosity, $\cs$ is the critical overlap concentration, $\etas$ is the viscosity of the solvent without polymer, $\eta$ is the shear viscosity of the polymer solution, and $\lambda_{\rm R}$ is the relaxation time.}
   \label{tbl:peoprop}
 \end{table*}
 
The experimental setup, shown in figure~\ref{fig:setup}(a), consists of slowly extruding a dilute polymer solution through a nozzle to form a pendant drop, which then detaches when its weight overcomes capillarity. As the drop falls, it is bound to the nozzle by a liquid neck that thins down, stretches into a long filament, and eventually breaks. Most of the experiments are performed using a nozzle of outer diameter $2.75\,\milli\meter$. Some additional experiments have been done with nozzles of outer diameter $\hnot= 0.4,\, 0.9,$ and $1.6\,\milli\meter$ (see Supplemental Materials). The quasistatic growth of the drop is achieved with a syringe pump (KDS Legato 110) at $Q=0.2\,{\rm mL/min}$ (see Supplemental Materials). We ensured that slightly changing the flow rate did not influence the dynamics reported in this study (see figure 1 in supplementary material). The thinning is recorded using a high-speed camera (Phantom VEO710), a macro lens (Nikon Micro-NIKKOR 200mm f/4 AI-s) and a microscope lens (Mitutoyo X2). The setup is backlit using a LED panel (Phlox), and the resulting resolution is about 10 \micro\meter\per pixels. Depending on the composition of the polymer solution, the frame rate is varied between 500 fps and 10,000 fps to ensure a precise measurement of the instantaneous strain rate of the liquid neck (see Supplemental Materials). After the recording, we extract the contour of the liquid using ImageJ, and use a custom-made Python routine to extract the thickness of the neck at its thinnest point \hmin.

The polymer solutions are prepared by dissolving polyethylene oxide (PEO, Sigma-Aldrich) into a mixture of deionized (DI) water and glycerol (Sigma-Aldrich). We use PEO of two different molecular weights: $M_{\rm w}=300 \,\kilo\gram\per\mole$ and $M_{\rm w}=4000 \,\kilo\gram\per\mole$, further referred to as 300K PEO and 4000K PEO. The mass concentration of PEO in the solution varies from $c=0.01\%$ to $c=1\%$. Varying the amount of glycerol (from 0 to 60\% per weight) enables us to tune the solvent viscosity (from $\etas=1\,\milli\pascal\usk\second$ to $\etas=10\,\milli\pascal\usk\second$). Water and glycerol are first mixed together before adding the PEO powder. The mixture is then placed on a roller mixer and gently mixed for at least 24 hours.

The surface tension of the solution is around $60\, \milli\newton\per\meter$ and does not vary significantly in the range of glycerol content and PEO concentration used here.~\cite{Christanti2001, Gaillard2021} We measured the viscosity of each solution using a rheometer (Anton Paar MCR 92) with a 50 mm-wide $1^{\rm o}$ cone-plate geometry. Figure~\ref{fig:setup}(b) shows the typical variations of the viscosity $\eta$ with the shear rate $\dot{\gamma}$ and with the polymer concentration of 300K PEO for a 50/50 wt\% water/glycerol solvent (see also Supplemental Materials). The viscosity of the PEO solutions increases with the polymer concentration. The shear viscosity of the PEO solutions remains Newtonian at small concentrations but exhibits a slight shear-thinning at the highest concentration used here. We provide in table \ref{tbl:peoprop} the range of concentrations used scaled by the critical overlap concentration \cs, defined as $\cs = 0.77/\etai$,\cite{Graessley1980} where $\etai = \left[ {(\eta-\etas)}/{\etas} \right]/c$ is the intrinsic viscosity. For a solution of 300K PEO in a 50/50 wt\% water/glycerol mixture, $\etai = 0.307\,\meter\cubed\per\kilo\gram$, and $\cs=0.22\%$.


\section{Results} \label{sec:3}

\subsection{Newtonian and viscoelastic regimes}

\begin{figure*}[!h]
\centering
  \includegraphics[width=0.99\textwidth]{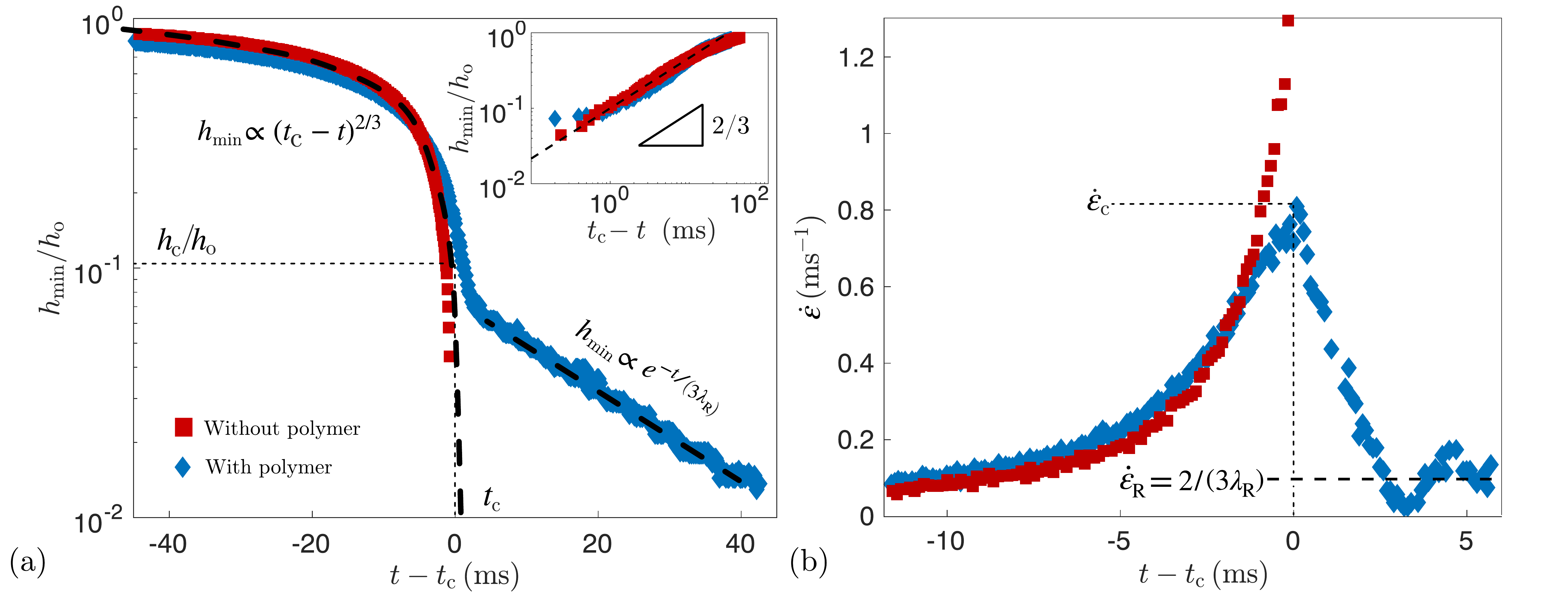}
  \caption{Time evolution of (a) the rescaled neck thickness $h_{\rm min}/h_{\rm o}$ and (b) the strain rate $\dot{\varepsilon}$ for the pure solvent (50/50 wt\% water-glycerol, red squares) and a dilute polymer solution ($c=0.5$\% 300K PEO in 50/50 wt\% water-glycerol, blue diamonds).
 The time is defined relative to the transition occurring at \tc.
      For $t<\tc$, the thinning is Newtonian and $h_{\rm min}$ follows the power law given by Eq.~(\ref{eq:newt}),
      with a similar dynamics for the solvent and the PEO solution.
      At $t=\tc$, the drop of solvent detaches ($h_{\rm min}=0$), but the PEO solution still has a finite thickness \hc.
      For $t>\tc$, the thinning of the PEO solution is viscoelastic, and the thickness decays exponentially 
      [Eq.~(\ref{eq:exp})].
      This regime corresponds to a constant strain rate $\dot{\varepsilon}_{\rm R}=2/(3\,\lambda_{\rm R})$.
       \SR{The inset in (a) shows the Newtonian regime on a logarithmic scale.}
  }
  \label{fig:regimes}
\end{figure*}

We first recall in figure~\ref{fig:regimes}(a) the classical difference between the thinning dynamics of a drop of dilute polymer solution (in blue) and a pure Newtonian solvent (in red). We observe that the dilute polymer solution follows the same dynamics as the solvent, up to a certain point. \SR{In the regime of small Ohnesorge numbers (${\rm Oh} = {\mu}/{\sqrt{\rho\sigma h_{\rm o}}} \lesssim 0.24$) considered here,} this first regime can be described as the thinning of an inviscid Newtonian fluid, which is driven by the capillary pressure, to which inertia resists. The thickness of the neck \hmin is given by the scaling law:~\cite{Kellert1983} 
\SR{
\begin{equation}
    \hmin = A(\eta)\left[\gamma(\tc-t)^2/\rho\right]^{1/3},
    \label{eq:newt}
\end{equation}
where $A(\eta)$ is a prefactor that weakly depends on the viscosity of the solvent, as reported by Thi\'evenaz \textit{et al.},\cite{Thievenaz2021}} $\gamma$ and $\rho$ are the surface tension and the density of the liquid, respectively. We can fit Eq.~(\ref{eq:newt}) to both the solvent and dilute polymer solution in the Newtonian regime, and obtain a good agreement as shown in figure~\ref{fig:regimes}(a). \SR{We obtain this self-similar thinning for about two decades of time, as shown in the inset of figure~\ref{fig:regimes}(a)}. In the following, we define \tc as the fitting parameter of Eq.~(\ref{eq:newt}), applied to the thinning of dilute polymer solutions in the Newtonian regime. Therefore, for the polymer solution \tc does not represent the moment of break-up, but the moment where the neck should have broken up if there were no polymer in the solution. Therefore, \tc represents the moment of the transition from the Newtonian to the viscoelastic regime. Using this definition of \tc, we can define the critical thickness of the neck at transition as $\hc = \hmin(\tc)$. 

Polymer solutions do not break at $t=\tc$ because of the presence of polymers. Initially coiled, the polymer chains eventually unwind under the increasing stress. This is the coil-stretch transition.~\cite{DeGennes1974, Christanti2001, Amarouchene2001} Once the polymer chains are unwound, they increase the viscosity of the solution and make it viscoelastic. As a result, for $t>\tc$, the thinning dynamics of the polymer solution slows down dramatically and enters the viscoelastic regime. The neck becomes a long filament of uniform thickness, which thins down exponentially:\cite{Anna2001}
\begin{equation}
    \hmin \propto e^{-t/(3\,\lr)},
    \label{eq:exp}
\end{equation}
where \lr is the longest relaxation time of the polymer. It has also been recently shown that the liquid interface exhibits a self-similar shape during the viscoelastic thinning of a polymer solution\cite{Deblais2020}. 

The viscoelastic thinning of dilute polymer solutions has been considered in many past studies.\cite{mckinley2002filament} In particular, it has been shown that by adding a dilute amount of polymer to a solvent, the dynamics can be described in terms of the thickness of the neck \hmin, or in terms of strain rate at the neck  $\strate = (\partial v_z/ \partial z)|_{h_{\rm min}}$, where $v_z$ is the axial component of the velocity.~\cite{Amarouchene2001} The latter can be expressed through the continuity equation:
\begin{equation}
    \strate = -\frac{2}{\hmin}\frac{ \partial \hmin}{ \partial t}.
\label{eqn:4}
\end{equation}
For a Newtonian fluid, the minimum neck thickness \hmin vanishes at time \tc, which is the moment the neck breaks. The evolution of \strate \,for both the water/glycerol mixture and the dilute solution of polymers is shown in figure~\ref{fig:regimes}(b). In terms of strain rate, the Newtonian regime of thinning given by Eq.~(\ref{eq:newt}) provides the expression:
\begin{equation}
    \strate = -\frac{4}{3} \frac{1}{\tc-t}.
    \label{eq:newt_strain}
\end{equation}
Therefore, for an inviscid Newtonian fluid, the strain rate diverges at pinch-off. For a dilute polymer solution, \tc also corresponds to the moment when the strain rate is maximal and reaches the critical strain rate $\stratec = \strate(\tc)$ before relaxing to a constant value. This constant strain rate in the viscoelastic regime is $\strater = 2/(3\lr)$.

\subsection{Influence of the polymer concentration and solvent viscosity on the thinning dynamics}

\begin{figure*}[!h]
\centering
  \includegraphics[width=0.99\textwidth]{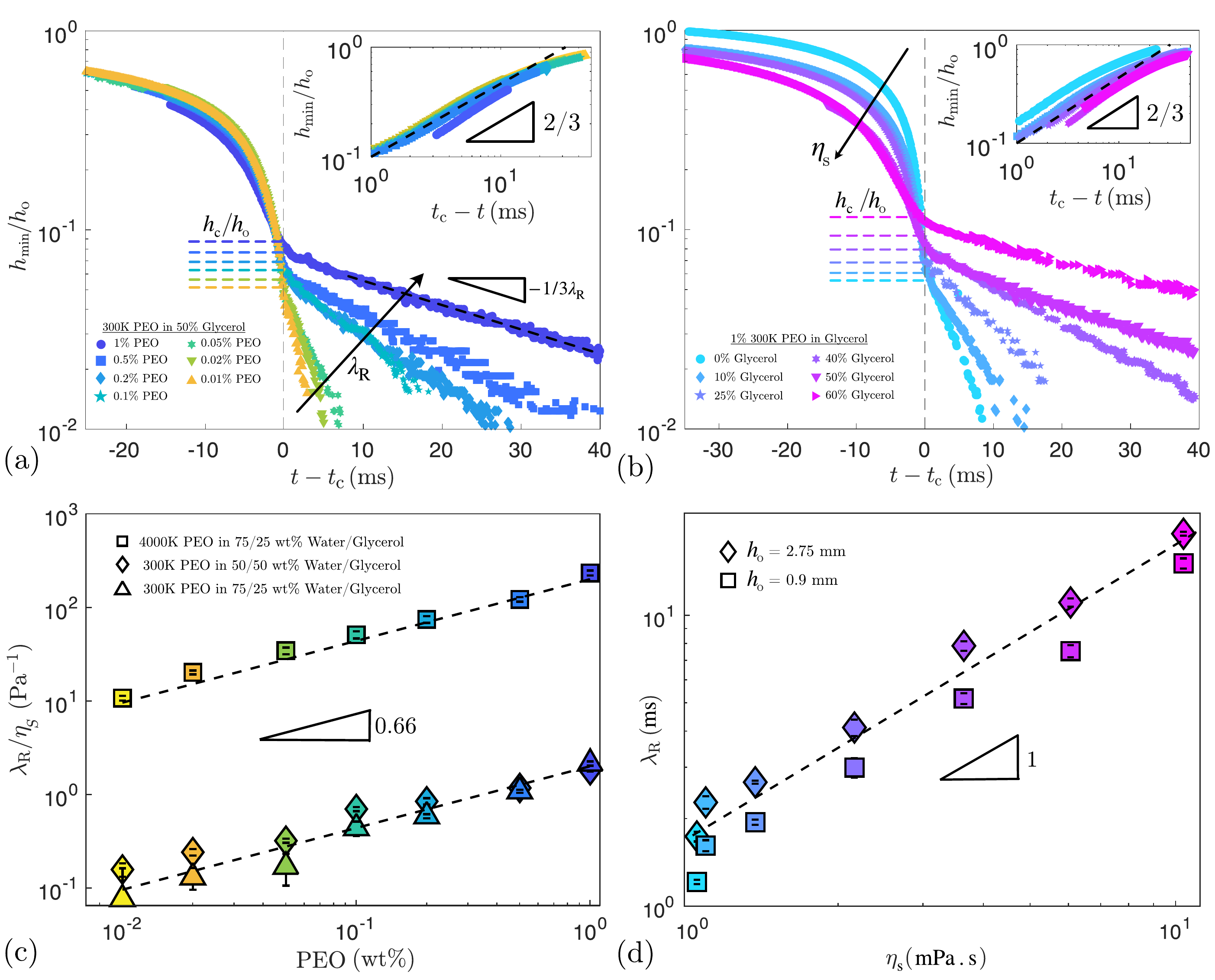}
  \caption{ 
      (a) Thinning dynamics for solutions of 300K PEO at various concentrations 
      (from yellow: $c=0.01$\% to dark blue $c=1\%$) in a 50/50 wt\% water/glycerol solvent.
      (b) Thinning dynamics for solutions with $c=1$\% of 300K PEO in solvents with different glycerol contents,
      hence different viscosity 
      (from light blue: water only, $\etas=\unit{1}\milli\pascal\usk\second$; to 
      pink: 40/60 wt\% water/glycerol, $\etas = \unit{10.41}\milli\pascal\usk\second$). The data shown corresponds to $\hnot = 2.75 \milli\meter$. \SR{Insets: Newtonian regime in log scale.}
      (c) Relaxation time in the viscoelastic regime \lr, rescaled by the solvent viscosity \etas, for increasing 
      polymer concentration.
      The dashed lines represent the empirical law $\lr/\etas \propto c^{0.66}$,~\cite{Tirtaatmadja2006}
      with a different prefactor depending on the molecular weight of the polymer.
      (d) Relaxation time \lr as a function of the solvent viscosity \etas for two nozzle sizes. \SR{The same samples are used for $h_{\rm o} = 0.9\,\milli\meter$ and $h_{\rm o} = 2.75\,\milli\meter$: solutions of 1\% 300K PEO and various water glycerol compositions.} The dashed line represents the linear fit given by Eq. (\ref{eq:zimm}).}
  \label{fig:thinning}
\end{figure*}

The different features of the two thinning regimes, Newtonian and viscoelastic, and of the transition between them depend on the molecular weight of the polymer, its concentration, and on the viscosity of the solvent. In figure~\ref{fig:thinning}(a), we show the thinning dynamics of solutions of 300K PEO, with mass concentrations varying between 0.01\% (yellow) and 1\% (dark blue), in a 50/50 wt\% water/glycerol solvent. First, we observe no significant influence of the polymer concentration on the Newtonian regime \SR{(inset of figure~\ref{fig:thinning}(a))}. Indeed, all the solutions follow the capillary-inertial scaling law given by Eq.~(\ref{eq:newt}). The small differences observed in this regime are due to the variations in shear viscosity reported in figure \ref{fig:setup}(b). The differences appear at the transition, \textit{i.e.}, at $t=\tc$. Depending on the concentration of polymers, the critical thickness of the neck at the transition, \hc, is different. Indeed, we observe that the more polymer in the solution, the thicker \hc is. We shall consider the variations of \hc with the different experimental parameters later in the article. In the viscoelastic regime, all dilute polymer solutions follow the exponential decay law given by Eq.~(\ref{eq:exp}), but with a different relaxation time \lr. As expected, \lr increases with the polymer concentration.\cite{del2017relaxation}

Similar to the effect of the polymer concentration, we can characterize the effect of the solvent viscosity. Figure~\ref{fig:thinning}(b) shows the thinning dynamics of solutions of 300K PEO with a mass concentration of 1\%, in water/glycerol mixtures where the mass fraction of glycerol is varied between 0\% (light blue, pure DI water) to 60\% (pink). At these moderate viscosities, the Newtonian regime still follows the capillary-inertial scaling law given by Eq.~(\ref{eq:newt}). Nevertheless, as expected, a higher viscosity yields a slower thinning. The effect of the viscosity, in this case, is simply to change the prefactor in Eq.~(\ref{eq:newt}).\cite{ThievenazSuspensions} Similar to increasing the polymer concentration, increasing the solvent viscosity also increases the critical thickness at the transition \hc, as well as the relaxation time in the viscoelastic regime \lr.

The relaxation time \lr is of great interest to predict the thinning of dilute polymer solutions and has been the focus of many studies.~\cite{Tirtaatmadja2006,Rodd2005} In figure~\ref{fig:thinning}(c), we report the quantity $\lr/\etas$ when varying the polymer concentration $c$, and recover the result of Tirtaatmadja~\textit{et al.}:~\cite{Tirtaatmadja2006} $\lr / \etas \propto c^{0.66}$. The relaxation time \lr is also expected to scale like the longest relaxation time \lz expected from the Zimm theory:\cite{Zimm1956,Tirtaatmadja2006}
\begin{equation}
    \lz \approx \frac{\etas \mw}{\na \kb T},
    \label{eq:zimm}
\end{equation}
where \na is the Avogadro number, \kb is the Boltzmann constant, and $T$ is the temperature. In figure~\ref{fig:thinning}(d), we also recover this linear trend between \lr and the solvent viscosity \etas.

This section was a reminder of classical results that aimed only at presenting the thinning dynamics, introducing the quantity of interest, and ensuring that we recover the results of the literature. The novelty in the analysis is presented in the following discussion, where we focus on the evolution of the strain rate at the transition. In particular, our results demonstrate that the transition also follows a self-similar evolution, whose only scale is the critical strain rate \stratec.

\begin{figure*}[!h]
\centering
  \includegraphics[width=0.98\textwidth]{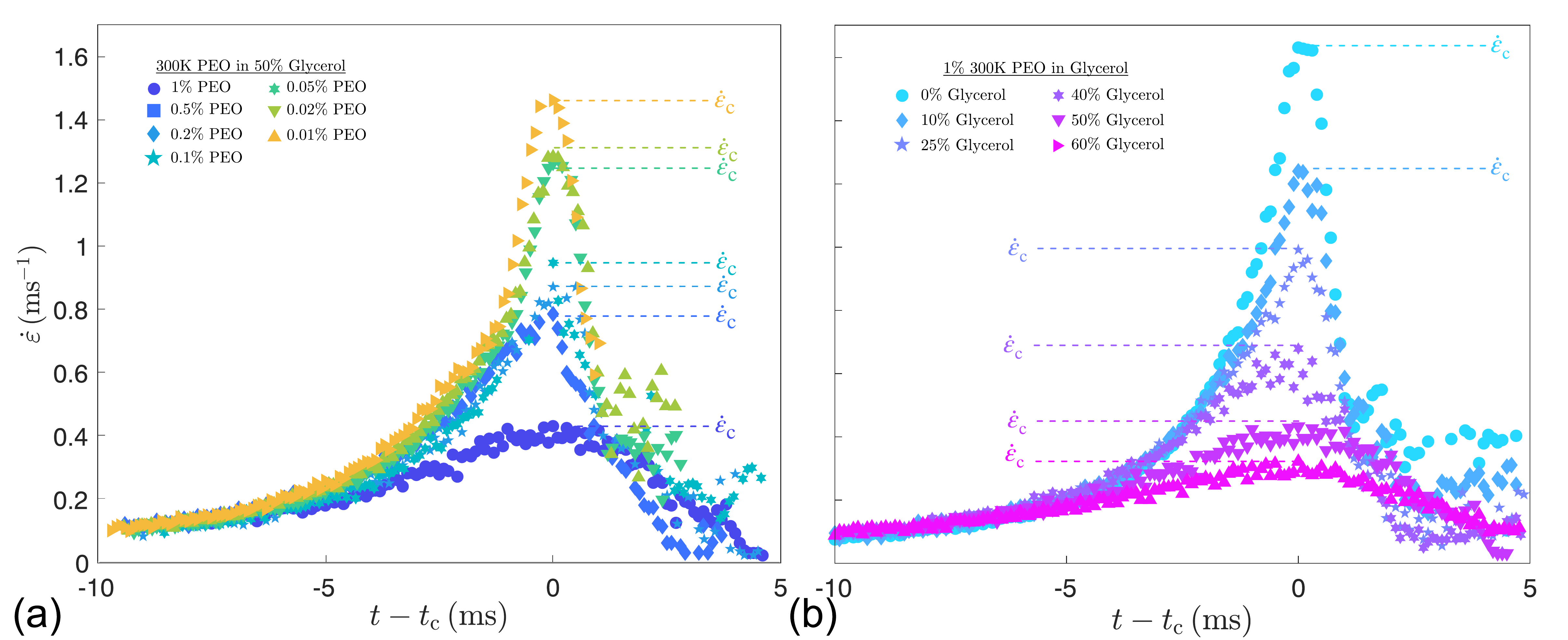}
  \caption{Time-evolution of the strain rate $\strate$ for dilute polymer solutions with 
  (a) different concentrations of 300K PEO in 50/50 wt\% water/glycerol solvent and 
  (b) $c=1\%$ 300K PEO in solvents with various water/glycerol mixtures.
  At the transition between the Newtonian and the viscoelastic regime ($t=t_{\rm c}$) \strate\,reaches it maximal value \stratec.}
  \label{fig:strate}
\end{figure*}


\begin{figure*}[!h]
\centering
  \includegraphics[width=0.97\textwidth]{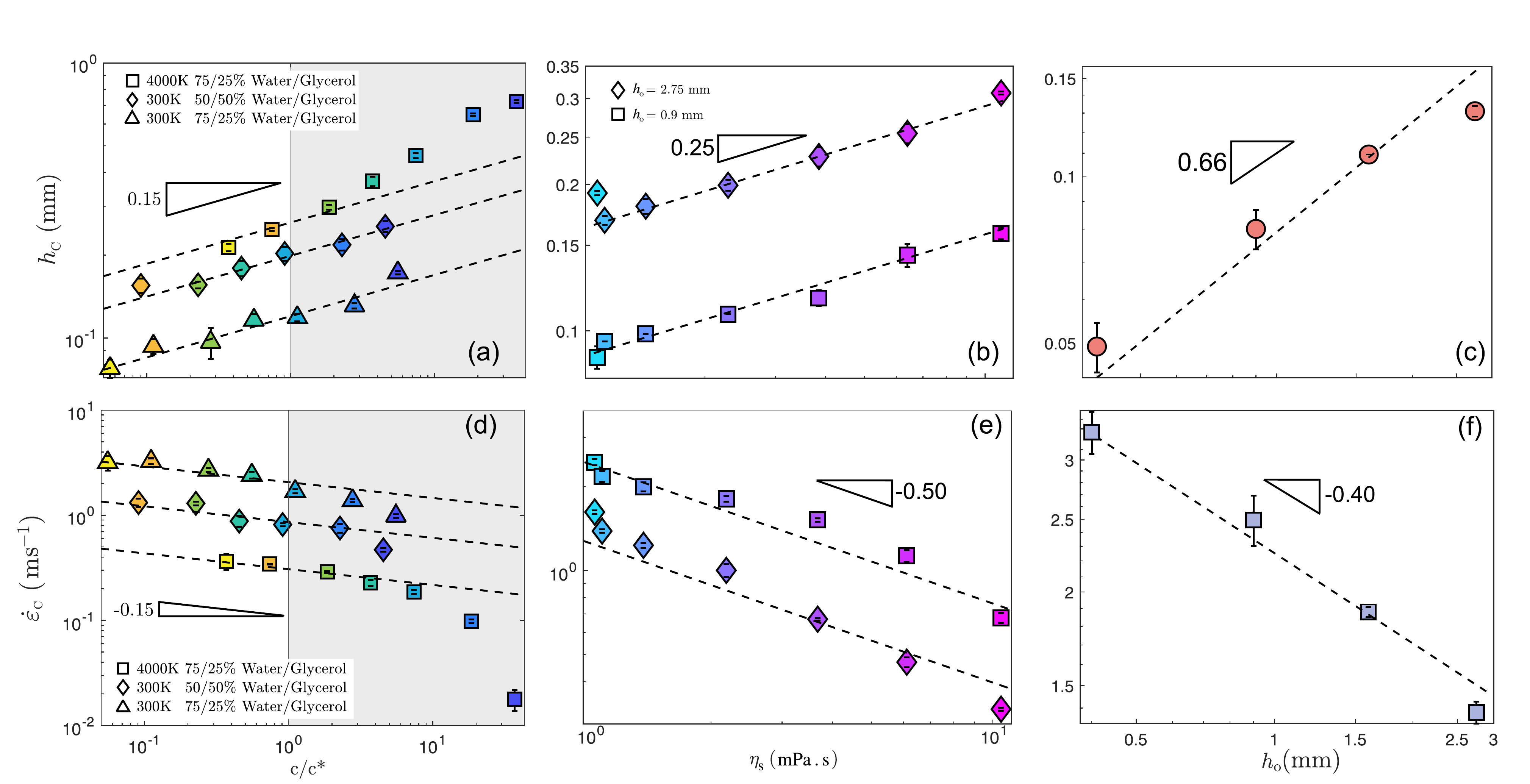}
  \caption{Evolution of (a)-(c) the critical neck thickness $h_{\rm c}$ and (d)-(f) the critical strain rate, $\dot{\varepsilon_{\rm c}}$ at the transition when varying (a) and (d) the \SR{non-dimensional concentration $c/\cs$} of the polymer for 300K and 4000K polymers in different solvents,  (b) and (e) the viscosity of the solvent \etas by varying the water/glycerol ratio for $c=1\%$ of 300K PEO, (c) and (f) the diameter of the nozzle \hnot for $c=0.5\%$ of 300K PEO in a 75/25 wt\% water/glycerol mixture.}
  \label{fig:critical}
\end{figure*}

As previously mentioned, the thinning of the polymer solution can also be described in terms of strain rate at the neck \strate, defined in Eq.~(\ref{eqn:4}). As illustrated in figure \ref{fig:regimes}(b), \strate \,increases sharply during the Newtonian regime, reaches a maximum \stratec at the onset of the coil-stretch transition, and immediately after quickly decreases. Figures~\ref{fig:strate}(a)-(b) show the evolution of $\strate(t)$ for the experiments reported in figures~\ref{fig:thinning}(a)-(b). When the polymer concentration increases (Figure~\ref{fig:strate}(a), from yellow to blue), the curve $\strate(t)$ flattens. A decrease in \stratec implies the transition lasts longer. The trend is also similar when the solvent viscosity increases (Figure ~\ref{fig:strate}(b), from light blue to pink). This similarity between the effects of polymer concentration and solvent viscosity suggests that a single mechanism drives the evolution of \stratec at the transition.

\subsection{Transition between the Newtonian and the viscoelastic regime}

The critical thickness \hc and the critical strain rate \stratec characterize the onset of the coil-stretch transition in the neck. These two parameters depend on all the experimental variables: the polymer concentration, the solvent viscosity, the molecular weight of the polymer, and the size of the drop. The evolution of \hc and \stratec with those parameters can be fitted with power laws, as reported in figures~\ref{fig:critical}(a)-(f).

As the polymer concentration increases, we observe an increase in the critical thickness \hc and a decrease in the critical strain rate \stratec as reported in figures~\ref{fig:critical}(a) and \ref{fig:critical}(d), respectively. Both quantities follow opposite power laws: $\stratec \propto c^{-0.15}$ and $\hc \propto c^{0.15}$. The prefactors of these scaling laws depend on the solvent viscosity and on the molecular weight of the polymer. The more viscous the solvent and the higher the molecular weight, the lower \stratec and the larger \hc. Therefore, increasing the viscoelastic behavior of the dilute polymer solution smooths the transition. For a high concentration of PEO (4000K, squares), the data match the scaling laws up to $c=0.2\%$, which is about ten times larger than the overlap concentration ($\cs = 0.027\%$). Therefore, it appears that the present scaling law for \hc and \stratec are valid for dilute and slightly semi-dilute solutions only.

Similarly, figures~\ref{fig:critical}(b) and \ref{fig:critical}(e) show that as the viscosity of the solvent increases, \stratec decreases and \hc increases, following the power laws $\stratec \propto {\etas}^{-0.50}$ and $\hc \propto {\etas}^{0.25}$. These scaling laws also depend on the diameter of the nozzle \hnot as demonstrated in figures~\ref{fig:critical}(c) and \ref{fig:critical}(f). We naturally recover the same trend: a wider nozzle leads to a lower critical strain rate \stratec and to a larger critical thickness \hc. We obtain the following scaling laws: $\stratec \propto {\hnot}^{-0.40}$ and $\hc \propto {\hnot}^{0.66}$. Finally, we can sum up the different contributions of these scaling laws:
\begin{equation}
    \stratec \propto c^{-0.15} \,{\etas}^{-0.50} \,{\hnot}^{-0.40};
    \quad
    \hc \propto c^{0.15}\, {\etas}^{0.25}\, {\hnot}^{0.66}.
    \label{eq:summary}
\end{equation}

\section{Discussion} \label{sec:4}

\subsection{Characteristic thinning rate of the neck during the transition}

\begin{figure}[t]
\centering
  \includegraphics[width=0.5\textwidth]{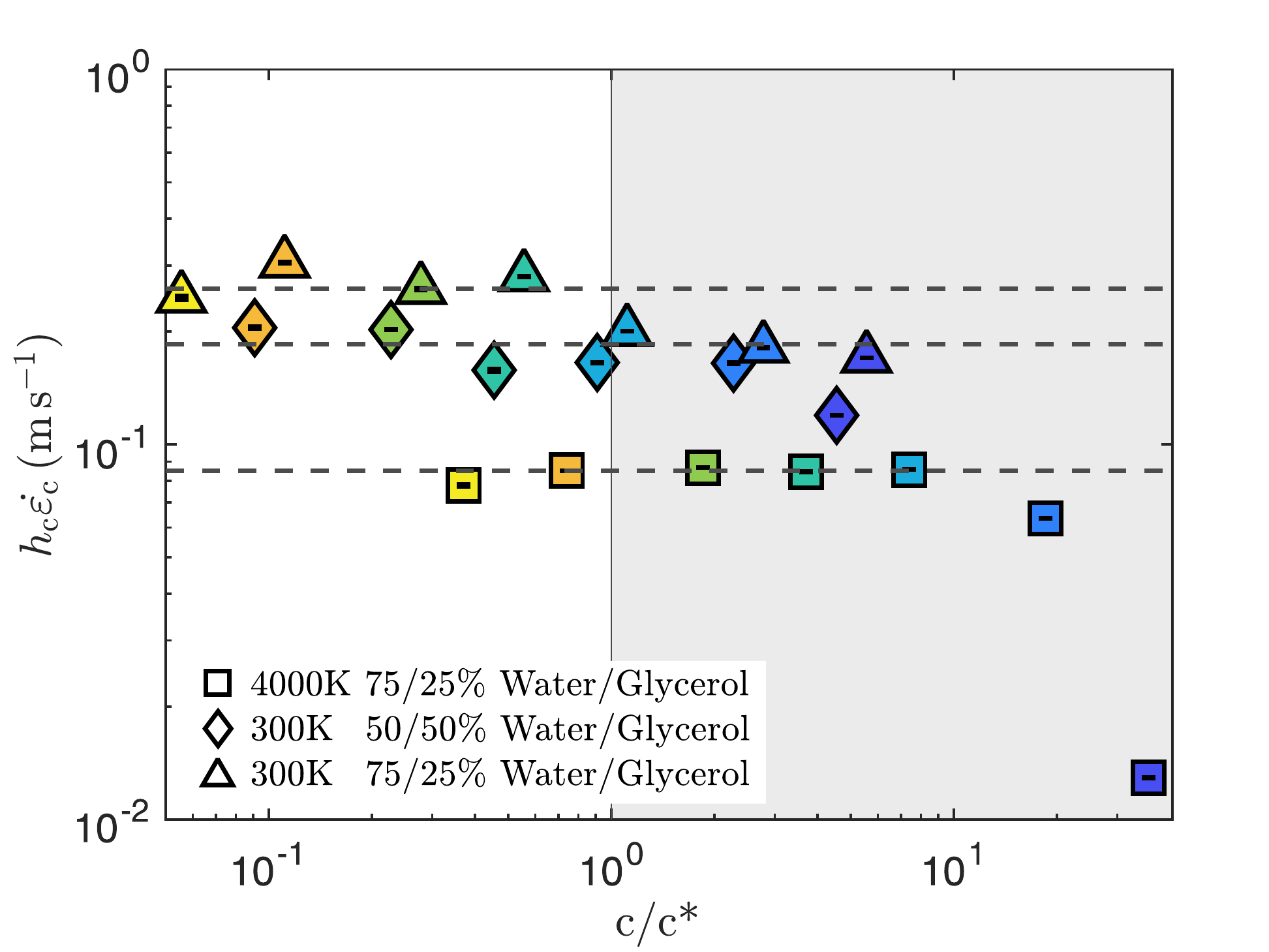}
  \caption{Evolution of $h_{\rm c} \dot{\varepsilon_c}$ with the \SR{non-dimensional} concentration of PEO for different solvent and molecular weight of the polymer. The horizontal dotted line shows constant values of $\vc$.}
  \label{fig:velocity}
\end{figure}

The characterization reported in figures~\ref{fig:critical}(a)-(f), and summed up in Eq.~(\ref{eq:summary}), suggests that the product $\hc \stratec$ does not depend on the polymer concentration. We also notice that this quantity has the dimension of a velocity, and we therefore define this parameter as a critical velocity $\vc = \hc \stratec$ of the neck during the transition. Figure~\ref{fig:velocity} plots the evolution of \vc when varying the \SR{non-dimensional} polymer concentration, for different viscosity of the solvent and polymer molecular weight. As expected, \vc is independent of the polymer concentration in the dilute regime. We observe the same abrupt decrease in the case of the 4000K PEO when the concentration $c$ crosses 0.2\% ($c = 7.4 \cs$), which is consistent with the limit of the dilute regime. The value of \vc depends on the molecular weight of the polymer. Indeed, we measure for the 300K PEO $\vc = \unit{0.26}\meter\per\second$ in 75/25 wt\% water/glycerol ($\etas=2.16\,{\rm mPa\,s}$) whereas we obtain $\vc = \unit{0.08}\meter\per\second$ for 4000K PEO. Note that according to Eq.~(\ref{eq:summary}), \vc also has a weak dependence in the solvent viscosity, following $\etas^{-0.25}$. This dependence is recovered when comparing the values obtained in the 50/50 wt\% water/glycerol ($\etas=6.06\,{\rm mPa\,s}$) where $\vc = \unit{0.185}\meter\per\second$ and in 75/25 wt\% water/glycerol.

\subsection{Self-similarity of the strain rate}

\begin{figure}[!b]
\centering
  \includegraphics[width=0.5\textwidth]{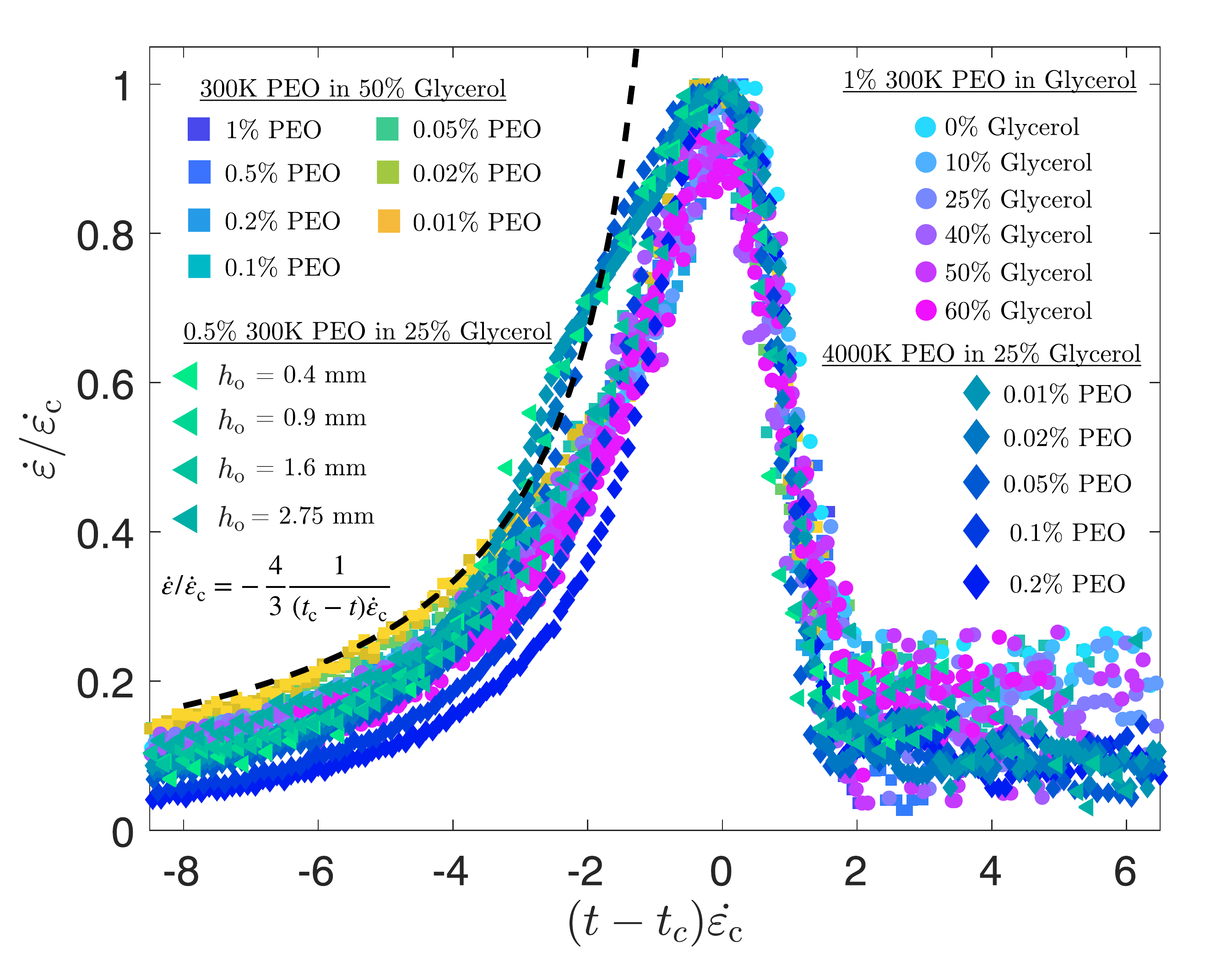}
  \caption{Evolution of the strain rate around the transition, rescaled by the critical strain rate: $\strate/\stratec$ as a function of $(\tc-t)\,\stratec$. The data plotted include different polymer concentrations ($c=0.01$\% to 1\%), 
  viscosity of the solvent ($\etas=1$ to 10\,\milli\pascal\usk\second), 
  molecular weights (300K and 4000K),
  and nozzle diameters ($\hnot=0.4$ to 2.75 \milli\meter).
  }
  \label{fig:self-similar}
\end{figure}

As previously mentioned in the description of figures~\ref{fig:strate}(a) and \ref{fig:strate}(b), the effect of the polymer concentration on the evolution of the instantaneous strain rate $\strate(t)$ is similar to the effect of the solvent viscosity \etas. Besides, the inverse of the critical strain rate, ${\stratec}^{-1}$, has the dimension of a time, so that we can use ${\stratec}^{-1}$ as a typical time scale of the transition.

Figure~\ref{fig:self-similar} shows the time evolution of the strain rate around the transition, rescaled by the time scale $\stratec^{-1}$, \textit{i.e.}, $\strate/\stratec$ as a function of $(t-\tc)\,\stratec$. In this figure, we included every parameter that we have investigated: the polymer concentration $c$, the molecular weight $M_{w}$, the solvent viscosity $\etas$, and the nozzle diameter $\hnot$. The dashed line represents the strain rate expected from Eq. (\ref{eq:newt_strain}) in the Newtonian regime; it follows the right trend despite a time shift of a few \milli\second, due to the numerical processing. We obtain a good collapse of all the data onto a single master curve \SR{during the transition, \textit{i.e.}, in the vicinity of $t \sim t_{\rm c}$. This notable result} leads to two key findings. First, the transition to the viscoelastic regime follows a self-similar dynamic. Second, there is only one physical quantity associated with this transition, the time scale $\stratec^{-1}$. In particular, since the nozzle diameter $\hnot$ is the only macroscopic length scale in the system, the good collapse of all our data implies that there is no macroscopic length scale in the mechanism of the transition from the Newtonian to the viscoelastic regimes.

The rescaling by \stratec is even more universal. A previous study showed that adding solid particles to a given polymer solution with fixed polymer concentration, molecular weight, and solvent viscosity, leads to the same rescaling.\cite{ThievenazSuspensions} The presence of rigid particles only influenced the value of \stratec without changing the mechanism of the transition. The present study extends this result to the case of any dilute solution of polymer in a good solvent.

\subsection{Interpretation of the critical strain rate}

We have shown in the previous section that the dynamics of the transition to the viscoelastic regime can be captured by the single parameter \stratec. Although we are not able to provide a theoretical model for the evolution of the critical strain rate with the different parameters, we can qualitatively explain its variations.

Two forces drive the thinning of the neck: the weight of the falling drop and the capillary pressure at the neck. Since the size of the falling drop is the result of the balance between its weight and the capillary force that held it to the nozzle, the weight of the drop is also linked to the capillary forces. As the neck deforms, the local viscous stress acting on the polymer chains increases until it becomes sufficient to unwind the chains.\cite{Amarouchene2001} We can therefore consider that the critical macroscopic strain rate \stratec is linked to the critical microscopic stress. The stronger the local stress, the more easily the polymer chains unwind. All things being equal, if the solvent is more viscous, then the viscous stress acting on each polymer chain will be stronger. If we use a wider nozzle, then the falling drop will be heavier and pull stronger on the neck, and so the local stress will also increase.

\begin{figure}[h]
\centering
  \includegraphics[width=0.5\textwidth]{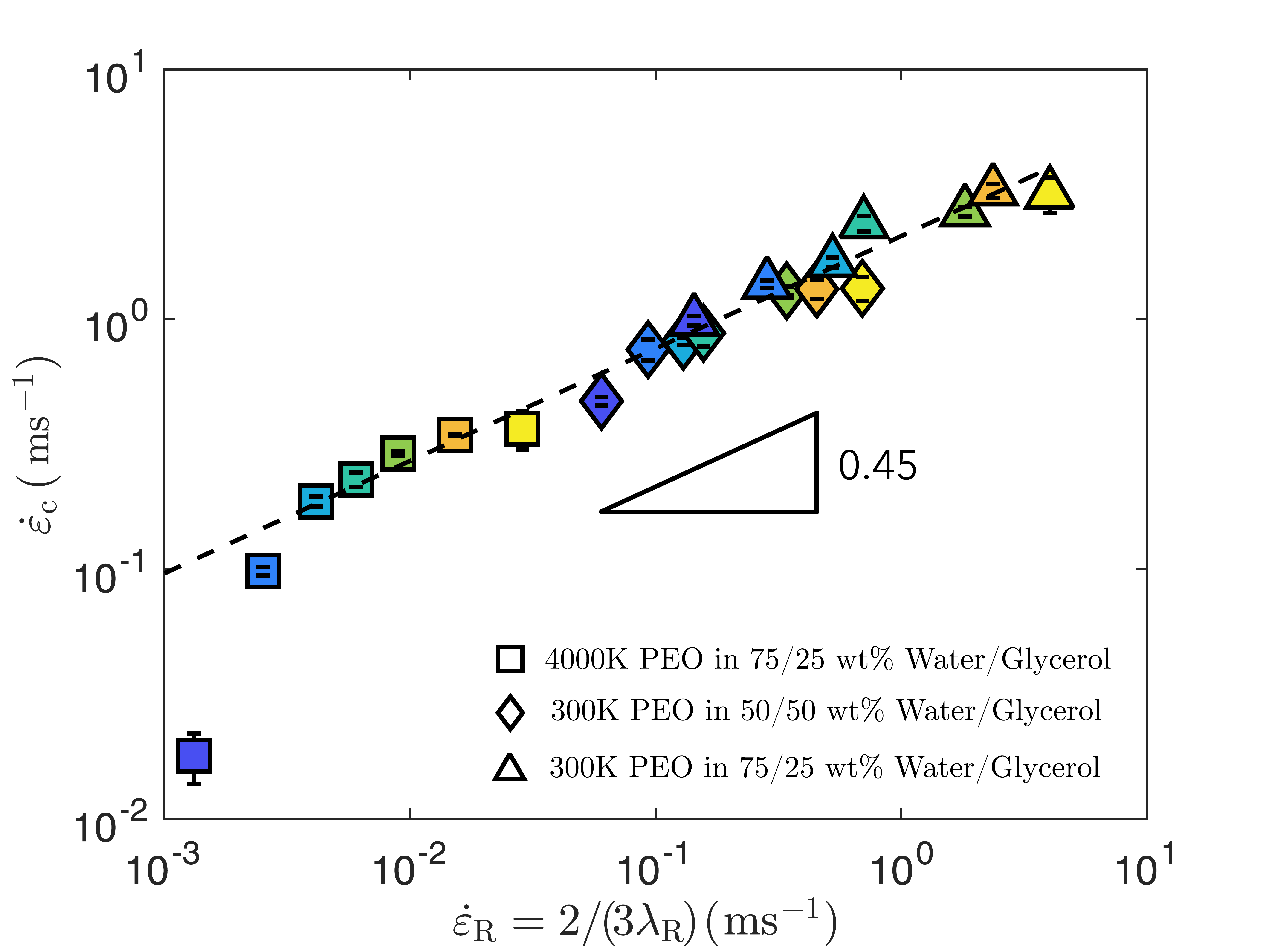}
  \caption{Critical strain rate \stratec as a function of the strain rate in the viscoelastic regime \strater.
  The dashed line represents the best power-law fit, $\stratec \propto \strater^{0.45}$, excluding the two leftmost points that are not in the dilute regime.}
  \label{fig:strain_relax}
\end{figure}

The effect of the polymer concentration is more complex. On the one hand, increasing the concentration increases the viscosity of the solution, and probably increases the local stress on the polymer chains. On the other hand, in the dilute regime, the viscosity of polymer solutions increases linearly with the concentration following Einstein's law.\cite{rubinstein2003polymer} If the sole effect of concentration on \stratec was through the linear increase of the viscosity, we should have similar exponents in the scaling laws given by Eq.~(\ref{eq:summary}). However, this is not the case: \stratec depends on the concentration in a much weaker way than it depends on the viscosity. Therefore, we can only conclude that in this dilute regime, the coil-stretch transition of a polymer chain is primarily controlled by the solvent viscosity, and secondarily by the concentration in polymers.

One can interpret 1/\stratec as the relaxation time of the polymer in the solvent: to stretch the polymer chain, one must deform it faster than it naturally relaxes. Figure~\ref{fig:strain_relax} compares the critical strain rate at the transition \stratec to the constant strain rate in the viscoelastic regime \strater. There are two main observations to be made. First, in most of our experiments, an order of magnitude separates \stratec and \strater. The strain rate at the transition \stratec is about ten times stronger than the strain rate in the viscoelastic regime \strater. Second, although \stratec and \strater vary in the same way, they do not follow a linear relationship. Instead, the best power-law fit yields $\stratec \propto \strater^{0.45}$. Note that this fit excludes the highest concentrations ($c>0.2\%$) for the 4000K PEO, which are not in the dilute regime.

This difference between the two relaxation times can be qualitatively explained by considering their reference states. \lr is the longest relaxation time of the polymer, and also the time scale of the stretching of the chain from the coiled state to the fully stretched state. On the other hand, $\stratec^{-1}$ is the time scale of the stretching from the coiled state, but not necessarily up to the fully stretched state. We can then consider that the abrupt decay of \strate \,corresponds to a partial stretching of the chains. At $t = \tc$, the chains are stretched just enough, so the viscosity of the solution rises significantly, inhibiting the liquid inertia. As the concentration, the molecular weight, or the solvent viscosity is increased, the relative stretching required to stop the thinning decreases: the polymer chains need to stretch less to cause the same change in the rheology of the solution.

This hypothesis is supported by figure~\ref{fig:strain_relax}. Indeed, for a low viscosity of the solvent (75/25 wt\% water-glycerol), a low polymer concentration ($c<0.05\%$) and a low molecular weight (300K PEO), \stratec and \strater have comparable values around 3\,\reciprocal{\milli\second}. Moreover, the actual break-up of the filament occurs shortly after the transition [see figure~\ref{fig:thinning}(a)]. This suggests that the chains fully stretch at the transition. On the other hand, for a more viscous solvent (50/50 wt\% water-glycerol), a higher polymer concentration ($c=1$\%) and higher molecular weight (4000K), \stratec is ten times larger than \strater. In this case, the chains are only partially stretched after the transition, and it takes much longer (about the longest relaxation time \lr) to fully stretch them.

\section{Conclusion}

In this study, we have characterized the transition between the Newtonian regime and the viscoelastic regime in the pinch-off of a drop of dilute polymer solution \SR{(here corresponding roughly to $c/c^*<5$)}. Macroscopically, this transition corresponds to the transformation of the liquid neck that binds the drop to the nozzle into a long filament. Microscopically, the transition corresponds to the coil-stretch transition of the polymer chains. Rather than the thickness of the neck, the relevant quantity to describe the thinning dynamics is the instantaneous strain rate \strate. We have shown that in terms of strain rate, the transition follows a universal self-similar dynamics that is only controlled by the critical strain rate \stratec. The quantity \stratec represents the ease with which the polymer chains unwind: the smaller \stratec, the less external deformation is required to trigger the transition. We have shown that a more viscous solvent, a higher molecular weight of the polymer, or a stronger external stress reduces \stratec and facilitates the transition to the viscoelastic regime. Moreover, it appears that a higher polymer concentration also facilitates the transition, independently of the viscosity increase that it causes. The typical time scale of the transition $\stratec^{-1}$, and therefore the time scale at which viscoelasticity appears, can be much shorter than the longest relaxation time of the polymer \lr. \SR{Different theoretical and numerical studies have used models like Oldroyd-B or FENE fluids, to successfully describe the formation of the ‘beads-on-a-string’ structure and the exponential thinning.\cite{li2003drop,Wagner2005,clasen2006beads,ardekani2010dynamics,turkoz2018axisymmetric} It would be interesting in future works to investigate if such an approach could successfully characterize the transition presented here}. A better understanding of the transition between the Newtonian and the viscoelastic regimes could be benficial to applications that require the break-up of ligaments of polymer solutions,\cite{chan2021torsional} and the atomization into droplets. \cite{Keshavarz2015,Keshavarz2016,Keshavarz2020,sen2021retraction}

\section*{Conflicts of interest}
There are no conflicts to declare.

\section*{Acknowledgements}
We gratefully acknowledge the discussions with Bavand Keshavarz.
This material is based upon work supported by the National Science Foundation under NSF CAREER Program Award CBET Grant No. 1944844.


\balance

\bibliography{transition_main.bbl} 
\bibliographystyle{transition_main} 

\end{document}


\maketitle

\vspace{5mm}

\noindent {\Large {\textsc{Influence of the flow rate on the thinning dynamics}}}

\vspace{5mm}

\begin{figure}[h!]
 \begin{center}
 \includegraphics[width=0.6\textwidth]{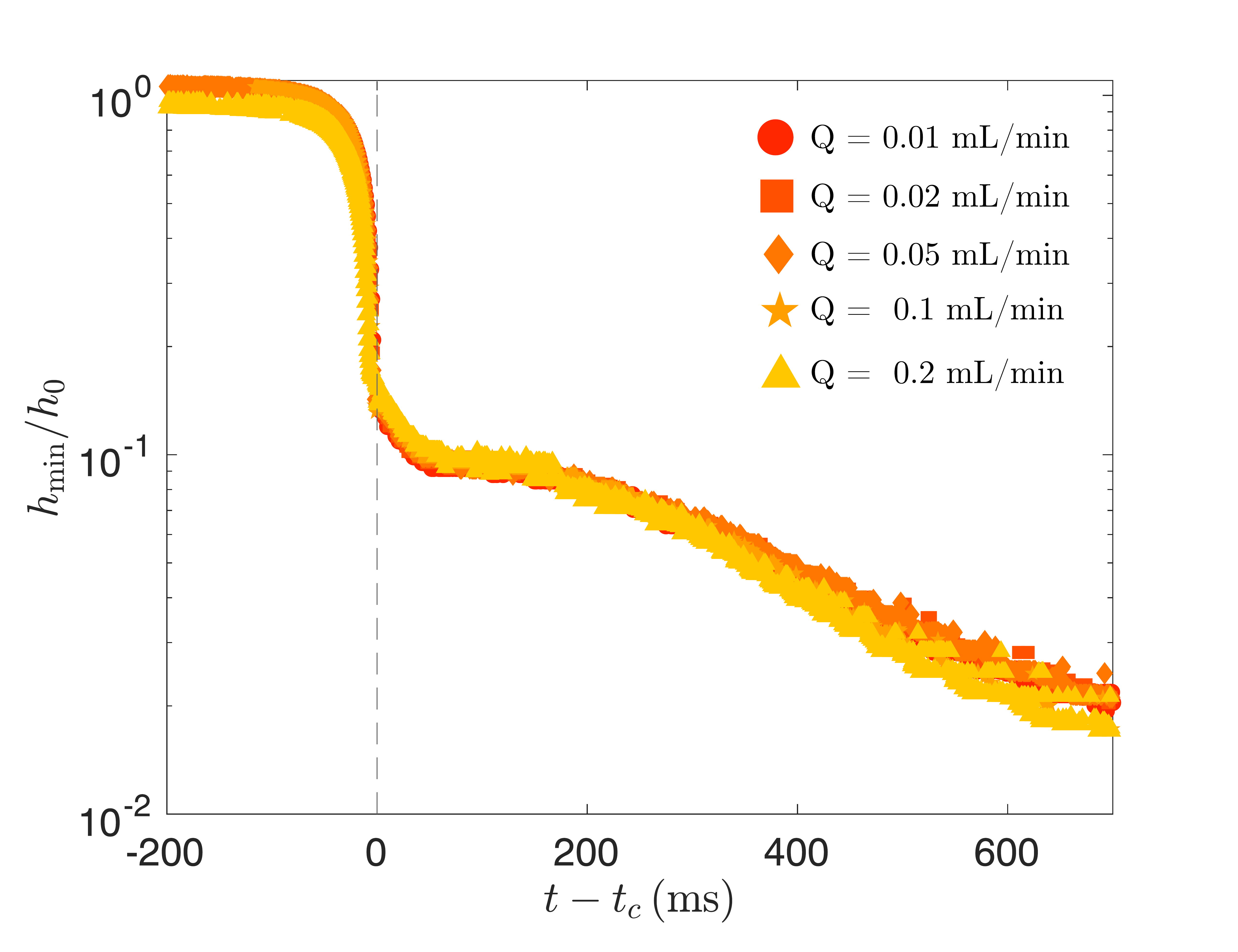}%
 \caption{Thinning dynamics of polymer solutions of 4000K PEO with $c = 0.5\%$ in a 75/25 wt\% water/glycerol solvent. The flow rates varies from $Q = 0.01$ to $0.2$ mL/min.}
 \label{fgr:SM_Fig1}
  \end{center}
 \end{figure}

\noindent To ensure that the thinning of the polymer solutions is not influenced by the flow rate set by the syringe pump, the pinch-off experiments were carried out at different flow rates. We show an example in figure \ref{fgr:SM_Fig1} for the thinning of a 4000K PEO of concentration $c = 0.5 \%$ in a 75/25 wt\% water/glycerol. The flow rates used are $Q = 0.01$, $0.02$, $0.05$, $0.1$, and $0.2$ mL/min. We observe that the thinning dynamics at those different flow rates collapse on the same master curve, thus ensuring that in this range of flow rate, the thinning occurs in a quasi-static regime.
\medskip

\vspace{10mm}

\newpage

\noindent {\Large {\textsc{Influence of the frame rate on the  critical strain rate}}}

\vspace{5mm}

\noindent For the experiments involving the 4000K PEO in 75/25 wt\% water/glycerol solvent, the frame rates were varied from 500 fps to 10000 fps depending on the concentration of polymer to account for the large range of relaxation times $\lambda_{\rm R}$ obtained. For example, we estimated the influence of the frame rate from 5000 fps to 50000 fps for the 4000K PEO at $c = 0.005\%$ in 75/25 wt\% water-glycerol solvent to ensure that the frame rate does not influence the measurement of $\dot{\varepsilon}_{\rm c}$ obtained at the transition. The concentration $c = 0.005\%$ was considered since this concentration led to the shortest relaxation time for this molecular weight. We also consistently increased the size of the window over which $\dot{\varepsilon}_{\rm c}$ was averaged with the frame rate. The mean strain rates over different frame rates is $\dot{\varepsilon}_{\rm c} = 1.1\, {\rm ms}^{-1}$ with less than 5\% deviation around this value.

\begin{figure}[h]
\centering
\includegraphics[width=0.6\textwidth]{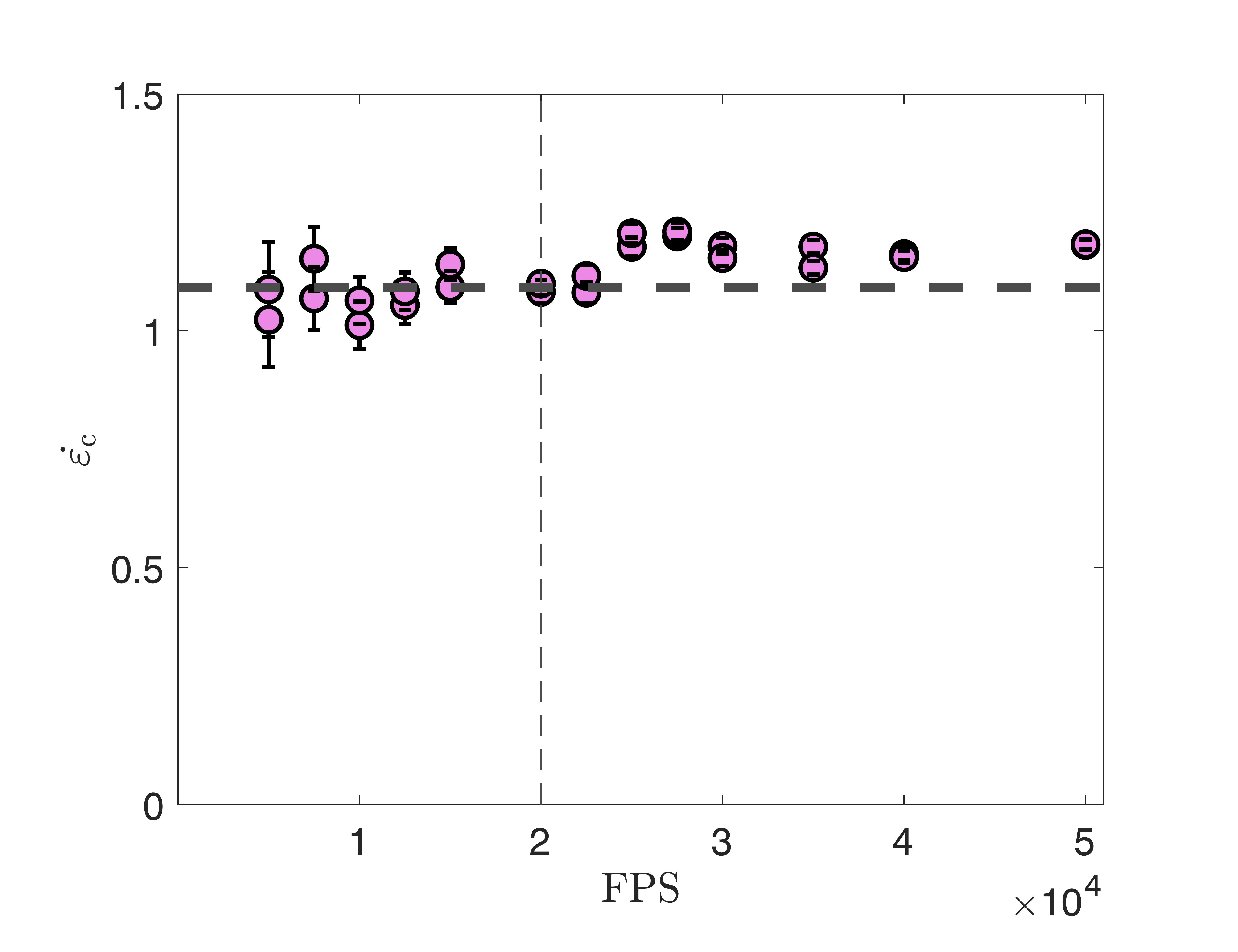}
   \caption{Evolution of $\dot{\varepsilon}_{\rm c}$ obtained experimentally when varying the frame rate per second (fps) for 4000K PEO at $c = 0.005\%$ in a 75/25 wt\% water-glycerol solvent. }
  \label{fgr:SM_Fig2}
\end{figure}

\newpage
\noindent {\Large {\textsc{Rheology of 300K and 4000K PEO in 75/25 wt\% water-glycerol solvent}}}

\vspace{5mm}

\noindent We measured the shear viscosity $\eta$ of the solution of 300K and 4000K PEO in a 75/25 wt\% water-glycerol solvent. We measured the viscosity of each solution using a rheometer (Anton Paar MCR 92) with a 50 mm-wide $1^{\rm o}$ cone-plate geometry. For the 300K PEO, the shear viscosity remains constant with the shear rate $\dot{\gamma}$ for $c = 0.01\%$ to $0.5\%$, and shows a weak shear-thinning at $c = 1\%$. The 4000K PEO the shear viscosity remains nearly constant between $c = 0.005\%$ and $0.02\%$, and then a shear-thinning is observed for $c = 0.05\%$ to $1\%$.

\begin{figure}[h]
\centering
\includegraphics[width=1\textwidth]{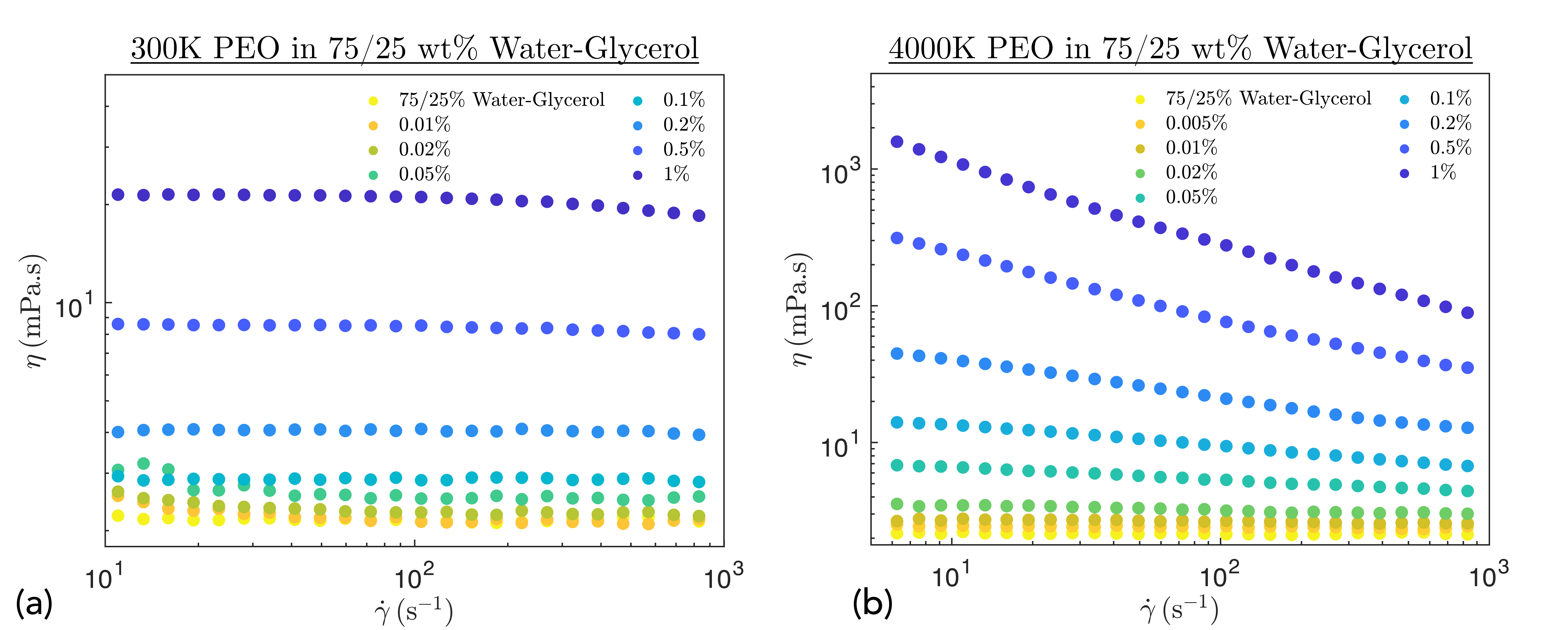}
   \caption{Evolution of the shear viscosity $\eta$ with the shear rate $\dot{\gamma}$ for polymer solutions prepared with different mass concentrations of (a) 300K PEO and (b) 4000K PEO in a 75/25 wt\% water-glycerol solvent.\label{fgr:SM_Fig3}}
\end{figure}

\newpage

\noindent {\Large {\textsc{Influence of the nozzle diameter on the thinning dynamics and the strain rate}}}

\vspace{5mm}

\begin{figure}[h]
\centering
\includegraphics[width=1\textwidth]{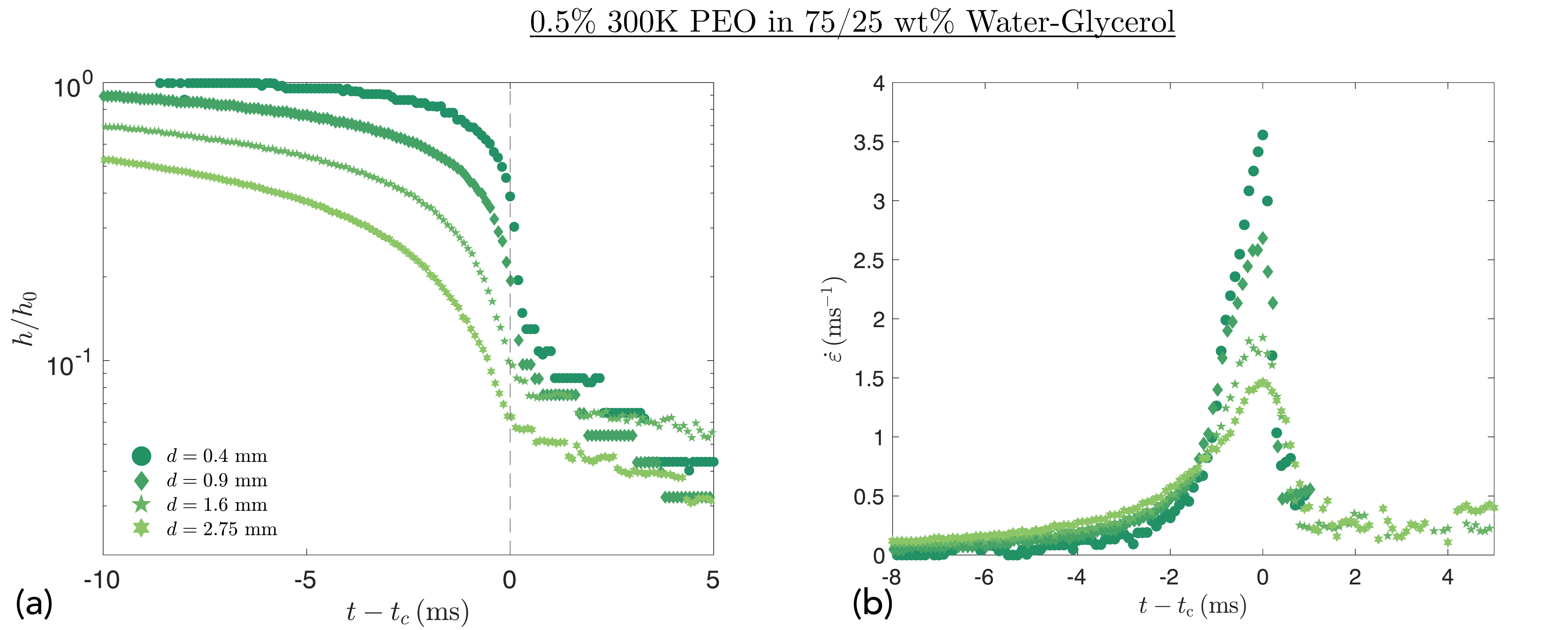}
   \caption{(a) Thinning dynamics $h(t)/h_0$ and (b) evolution of the strain rate $\dot{\varepsilon}(t)$ of a 300K PEO solution at $c = 0.5\%$ in 75/25 wt\% water-glycerol solvent for different nozzle diameters $h_{0} = 0.4, 0.9, 1.6\,\&\,2.75\,mm$.\label{fgr:SM_Fig4}} 
\end{figure}

\noindent To elucidate the influence of weight of the drop on the thinning and the pinch-off, we considered nozzles of different diameters $h_{0} = 0.4$, $0.9$, $1.6$, and $2.75$ mm. Here, we show the results for the 300K PEO at $c = 0.5\%$ in a 75/25 wt\% water-glycerol solvent. The corresponding thinning $h(t)$ and $\dot{\varepsilon}(t)$ are shown in figures \ref{fgr:SM_Fig4}(a) and \ref{fgr:SM_Fig4}(b), respectively. 

\newpage

\noindent {\Large {\textsc{Movies of the experiments.}}}
\bigskip

\noindent - Movie corresponding to the pinch-off of a droplet of 50/50 wt\% Water-Glycerol solvent at 10000 fps, shown in figure 1(a) and used in figures 3(a)-3(b) in the main manuscript.

\noindent - Movie corresponding to the pinch-off of a droplet of $c=0.5$\% 300K PEO in a 75/25 wt\% Water-Glycerol Solvent at 10000 fps, shown in figure 1(b) in the main manuscript.

\noindent - Movie corresponding to the pinch-off of a droplet of $c=0.5$\% 300K PEO in 50/50 wt\% Water-Glycerol Solvent at 10000 fps, used for the data shown in figures 3(a)-3(b) in the main manuscript.